# Demonstrating of Cosmic Ray Characteristics by Estimating the Cherenkov Light Lateral Distribution Function for Yakutsk Array as a Function of the Zenith Angle

Marwah M. Abdulsttar[a], A. A. Al-Rubaiee[b, *], Abdul Halim Kh. Ali[c]

College of Science, Dept. of Physics, Al-Mustansiriyah University,

10052 Baghdad, Iraq

[a]gonti77@mail.ru, [b]dr.ahmedrubaiee@gmail.com, [c]halimkh@yahoo.com

**Abstract**. Cherenkov light lateral distribution function (CLLDF) in Extensive Air Showers (EAS) for different primary particles ($e^-$, n , p, F, K and Fe) was simulated using CORSIKA code for conditions and configurations of Yakutsk EAS array with the fixed primary energy 3 PeV around the knee region at different zenith angles. Basing on the results of CLLDF numerical simulation, sets of approximated functions are reconstructed for different primary particles as a function of the zenith angle. A comparison of the parametrized CLLDF with that simulated with Yakutsk EAS array is verified. The parameterized CLLDF also is compared with that measured on the Yakutsk EAS array.

## 1. Introduction

Cosmic rays (CRs) are high-energy radiation arriving from the outside of the Earth's atmosphere. When an ultra-high energy CR particles enter the atmosphere they will eventually collide with an atmospheric nuclei. This interaction begins a cascade process in the atmosphere called EAS [1, 2]; where the density of the air shower is characterized by CLLDF. In the high and ultrahigh energy regions the only enable method of CR registration is indirect way from EAS induced in atmosphere, precisely by registration of atmospheric Cherenkov light. The investigation of CRs based on registration of Cherenkov light of secondary particles produced in the cascade processes of EAS has intensively been developed in the last years [3, 4]. One of the necessary tools of the numerical simulation is the Monte Carlo methods, which is used for investigating of EAS properties and experimental data analyzing [5]. In Refs. [6, 7] were performed the simulation of CLLDF using CORSIKA code for conditions of Yakutsk EAS array at the high energy range ($10^{13}$-$10^{16}$) eV for different primary particles (primary protons, Iron nuclei, Oxygen, Nitrogen, Calcium, Neutron and Argon) and different zenith angles ($0^o$, $5^o$, $10^o$ and $15^o$). By depending on the Breit-Wigner function, a parameterization of CLLDF was reconstructed by depending on CORSIKA simulation as a function of the primary energy.

In the present work, the simulation of CLLDF was performed for conditions and configurations of Yakutsk EAS array using CORSIKA code [8, 9] with the energy 3 PeV around the knee region for primary particles ($e^-$, n, p, F, K and Fe) at different zenith angles.The approximation with polynomial fit is obtained for the simulated results of Cherenkov light density was performed basing on Breit-Wigner functions [10, 11]. The evaluated parameters of CLLDF as a function of the zenith angle are compared with the simulated CLLDF using CORSIKA code and with the Yakutls experimental data.

## 2. Cherenkov Radiation

Cherenkov Light that created by EAS particles is produced when charged particle moving in some dielectric medium distorts the atoms in its local vicinity along its track and creating an electric polarization field. When a charged particle moves through dielectric medium and polarizes the medium when it comes close to its molecules that is within its Debye length so that like charges are displaced away in the atoms and the atoms behave like elementary dipoles. Fig.1 (a) shows at low velocity the atoms of the medium are undistorted and the electric field is symmetric both in azimuth and along the axis i.e. there is no resultant electric field and so no radiation. In Fig.1 (b) the atoms of the medium are at high velocity thus the polarization field produced by the moving particle is no longer symmetric along the axis of motion. In this case there is a resultant dipole field along the axis and the electric field will radiate a short electromagnetic pulse [2].

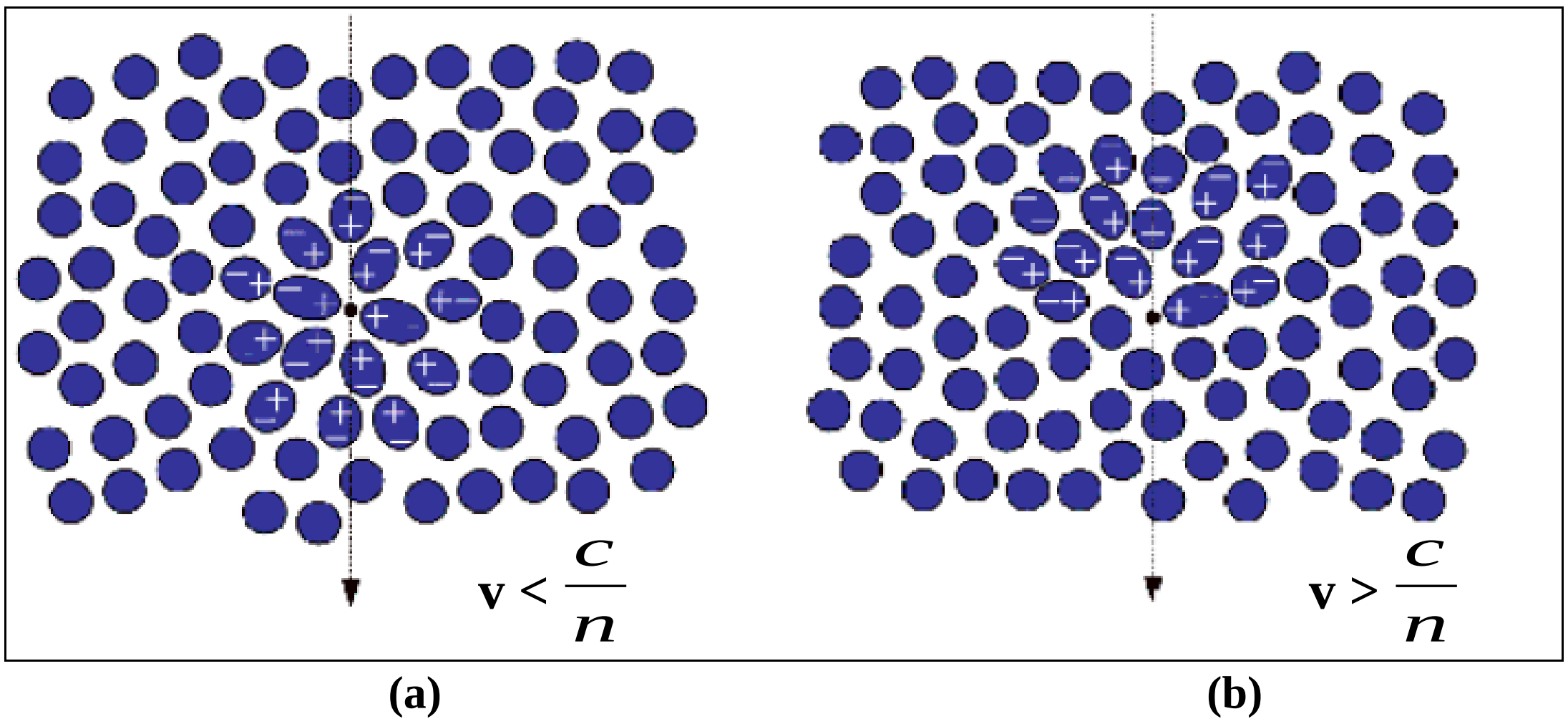

**Fig. 1** Charged particle moves through dielectric medium at (a) low velocity; (b) high velocity [2].

Charged particles passing a dielectric medium with a velocity faster than the velocity of light in that medium will emit radiation in cone like shape wave front behind them which is very like the shock wave generated by supersonic jets as shown in Fig.2 [12].

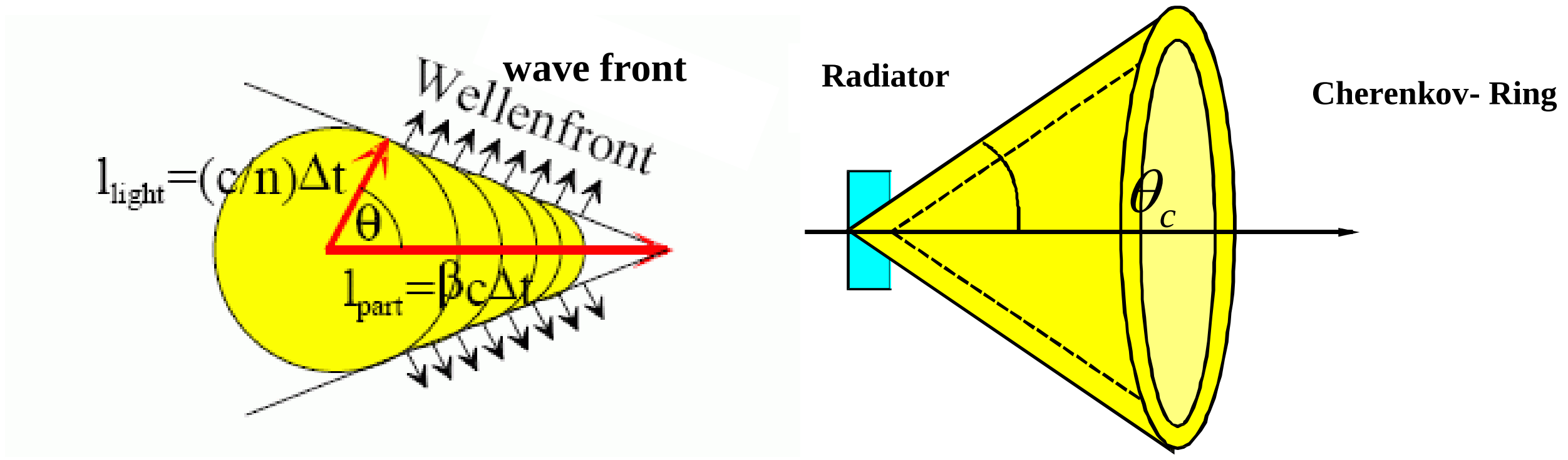

**Fig. 2** Emission of Cherenkov light by a charged particle traversing a medium with a velocity higher than the velocity of light in that medium [12].

The Cherenkov radiation is emitted into a cone which half opening angle $\theta_c$ is given by [2]:

$$\cos\theta = \frac{(c/n)t}{(\beta c)t} = \frac{1}{\beta n} \tag{1}$$

where $\beta=v/c$ ; $c$ is the velocity of light in free space and $n$ is the refractive index of the dielectric medium [13].

I. E. Tamm and I.M. Frank explained Cherenkov's experiment and derived a fundamental formula from their original work; the radiated energy per unit length [14, 15]:

$$\frac{dW}{dl} = \frac{e^2}{c^2}\int(1-\frac{1}{\beta^2 n^2}).\omega d\omega \tag{2}$$

The number of Cherenkov photons emitted by an electron at the wavelength interval ($\lambda_1$, $\lambda_2$) is given by [2]:

$$N = 2\pi\alpha l\left(\frac{1}{\lambda_2}-\frac{1}{\lambda_1}\right).\left(1-\frac{1}{\beta^2 n^2}\right) \tag{3}$$

From Eq. 1 and under the assumption that the refractive index is independent of the wave length, over a small frequency range we get [2]:

$$N = 2\pi\alpha l\left(\frac{1}{\lambda_2}-\frac{1}{\lambda_1}\right).\sin^2\theta_c \tag{4}$$

Where $\alpha$ is the fine structure constant = $e^2/\hbar c$ = 1/137

### 3. The Yakutsk EAS array

The Yakutsk array is located in Oktyomtsy near Yakutsk, Russia (61.7$^o$ N; 129.4$^o$ E), 100 m above sea level (1020 g/cm$^2$) and wavelength range from 300 to 600 nm. Its air Cherenkov light detectors are designed to detect CRs of extremely high energies i.e. the energy spectrum of CRs in the range $10^{15}$–$10^{19}$ eV. Now, the Yakutsk array consist 58 ground–base and six underground scintillation detector stations to detect (electrons and muons) and 48 detectors - PMTs in shuttered housing to detect the atmospheric Cherenkov light components of EAS[9]. The total area covered by Yakutsk array detectors is 10 km$^2$. In the central part of the array there is a denser domain with 100-250 m detector spacing. Through the whole period of observation approximately $10^6$ showers of the primary energy higher than $10^{15}$ eV are detected; the three highest energy events selected with axes within the array area and zenith angle ≤ 60$^o$ have an energy E > $10^{20}$ eV. Fig.3 shows the Yakutsk array detectors configurations where open circles represents charged particle detectors while filled circles are Cherenkov light detectors of the $C_1$ (~500m spacing) subset on the other hand, filled triangles for $C_2$ subset (50–200m spacing) the Cherenkov light detectors - open photomultiplier tubes (PMTs) of 176 cm$^2$ and 530 cm$^2$ acceptance area and squares indicate PMTs of autonomous sub-array with independent trigger which are the muon detectors [9].

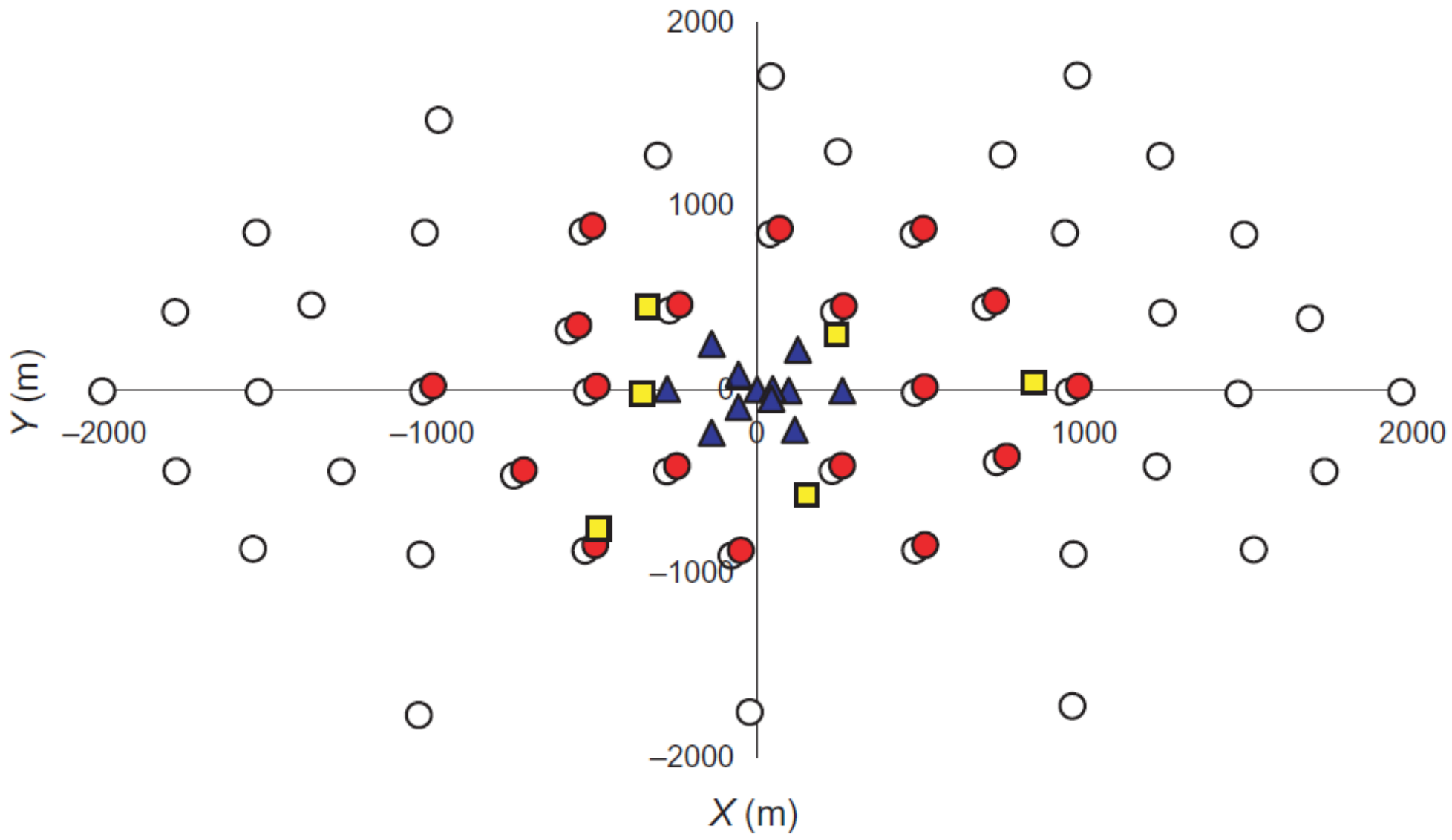

**Fig. 3** The Yakutsk EAS array detector configurations [9].

## 4. The simulation and parameterization of CLLDF

The simulation of CLLDF was performed by the computer code CORSIKA (Cosmic Ray Simulations for KAscade) which is a detailed Monte Carlo program to study the evolution and properties of EAS in the atmosphere [16] with using two hadronic models: QGSJET (Quark Gluon String model with JETs) code [17] that was used to model interactions of hadrons with energies higher than 80 GeV and GHEISHA (Gamma Hadron Electron Interaction SHower) code [18] which is used for energies lower than 80 GeV. The simulation of CORSIKA code was performed for conditions and configurations of Yakutsk EAS array with fixed energy 3PeV for different primary particles ($e^-$, n, p, F, K, and Fe) at different zenith angles ($5^o$, $10^o$, $15^o$, $20^o$, $25^o$) and azimuth angle $\Phi=0^o$ but for iron nuclei $\Phi$ taken from $0^o$ to $360^o$.

In order to parameterize the results of the simulated CLLDF, we used the suggested function in Ref. [3] which depends on four parameters a, b, σ and $r_o$. This function was normalized by the coefficient $C=10^3$ $m^{-1}$ [19]:

$$Q(\theta,R)=\frac{C\sigma\exp[a-G]}{b\left[(R/b)^2+(R-r_o)^2/b^2+R\sigma^2/b\right]} \tag{5}$$

Where $G$ is defined as:

$$G=R/b+(R-r_o)/b+(R/b)^2+(R-r_o)^2/b^2 \tag{6}$$

Where $R$ is the distance from the shower axis and $\theta$ is the zenith angle. The values of parameters $a$, $b$, $\sigma$ and $r_o$ were estimated by fitting the Eq. 5 to the values of the CLLDF that simulated by the CORSIKA code package with the energy 3 PeV around the knee region for different primary particles at different zenith angles. Unlike Refs. [19, 20], the parameters of CLLDF were parameterized as a function of the zenith angle $\theta$ using the third fit polynomial, which given by the relation:

$$K(\theta)=c_0+c_1\log_{10}(\theta)+c_2(\log_{10}\theta)^2+c_3(\log_{10}\theta)^3 \tag{7}$$

where $c_o$, $c_1$, $c_2$ and $c_3$ are coefficients depending on the type of the primary particles and the primary energy and their values are given in (Tables 1 and 2 ).

**Table 1:** Coefficients $c_i$ that determine zenith angle dependence (Eq. 7) of the parameters *a, b,* $\sigma$ and $r_o$ for primary electron, neutron, and proton for vertical showers of Yakutsk Cherenkov EAS array.

| $e^-$ | | | | | |
|---|---|---|---|---|---|
| $k$ | $c_0$ | $c_1$ | $c_2$ | $c_3$ | $\chi^2$ |
| $a$ | $10.617.10^{0}$ | $-95.423.10^{-2}$ | $10.671.10^{-1}$ | $-50.915.10^{-2}$ | $76.9.10^{-4}$ |
| $b$ | $55.369.10^{-2}$ | $-10.858.10^{-2}$ | $12.853.10^{-2}$ | $-65.550.10^{-3}$ | $76.9.10^{-4}$ |
| $\sigma$ | $27.131.10^{-1}$ | $-80.088.10^{-1}$ | $67.156.10^{-1}$ | $-18.115.10^{-1}$ | $23.10^{-3}$ |
| $r_o$ | $19.590.10^{-1}$ | $-73.231.10^{-1}$ | $57.904.10^{-1}$ | $-14.083.10^{-1}$ | $25.9.10^{-4}$ |
| n | | | | | |
| $a$ | $98.492.10^{-1}$ | $-14.639.10^{-2}$ | $55.867.10^{-2}$ | $-35.352.10^{-2}$ | $11.7.10^{-4}$ |
| $b$ | $41.834.10^{-2}$ | $93.610.10^{-3}$ | $-34.310.10^{-3}$ | $-13.94.10^{-3}$ | $13.3.10^{-4}$ |
| $\sigma$ | $-49.68.10^{-3}$ | $-79.349.10^{-2}$ | $61.468.10^{-2}$ | $-12.388.10^{-2}$ | $30.4.10^{-4}$ |
| $r_o$ | $-31.583.10^{-1}$ | $76.260.10^{-1}$ | $-81.049.10^{-1}$ | $27.802.10^{-1}$ | $14.10^{-5}$ |
| p | | | | | |
| $a$ | $10.132.10^{0}$ | $-56.516.10^{-2}$ | $67.262.10^{-2}$ | $-28.036.10^{-2}$ | $1.10^{-5}$ |
| $b$ | $-13.233.10^{-2}$ | $18.006.10^{-1}$ | $-16.781.10^{-1}$ | $50.035.10^{-2}$ | $31.10^{-5}$ |
| $\sigma$ | $-18.282.10^{-1}$ | $53.425.10^{-1}$ | $-55.149.10^{-1}$ | $18.555.10^{-1}$ | $16.9.10^{-3}$ |
| $r_o$ | $-37.606.10^{-1}$ | $96.560.10^{-1}$ | $-10.103.10^{0}$ | $34.489.10^{-1}$ | $4.10^{-5}$ |

**Table 2:** Coefficients $c_i$ that determine zenith angle dependence (Eq. 7) of the parameters *a, b,* $\sigma$ and $r_o$ for primary Fluorine, Potassium, and Iron nuclei for vertical showers of Yakutsk Cherenkov EAS array.

| F | | | | | |
|---|---|---|---|---|---|
| $k$ | $c_0$ | $c_1$ | $c_2$ | $c_3$ | $\chi^2$ |
| $a$ | $10.389.10^{0}$ | $-21.804.10^{-1}$ | $24.405.10^{-1}$ | $-87.080.10^{-2}$ | $4.10^{-5}$ |
| $B$ | $44.643.10^{-2}$ | $-38.670.10^{-3}$ | $58.380.10^{-3}$ | $-25.180.10^{-3}$ | 0 |
| $\sigma$ | $-19.025.10^{-1}$ | $47.184.10^{-1}$ | $-42.934.10^{-1}$ | $12.958.10^{-1}$ | $26.1.10^{-4}$ |
| $r_o$ | $-54.727.10^{-1}$ | $14.039.10^{0}$ | $-13.699.10^{0}$ | $43.882.10^{-1}$ | $62.10^{-5}$ |
| K | | | | | |
| $A$ | $10.086.10^{0}$ | $-95.882.10^{-2}$ | $10.039.10^{-1}$ | $-36.925.10^{-2}$ | $6.10^{-5}$ |
| $B$ | $49.507.10^{-2}$ | $-15.464.10^{-2}$ | $16.195.10^{-2}$ | $-60.10^{-3}$ | $2.10^{-5}$ |
| $\sigma$ | $-43.865.10^{-2}$ | $11.873.10^{-1}$ | $-21.094.10^{-1}$ | $98.899.10^{-2}$ | $27.5.10^{-3}$ |
| $r_o$ | $-71.530.10^{-2}$ | $12.795.10^{-1}$ | $-29.167.10^{-1}$ | $14.592.10^{-1}$ | $92.3.10^{-4}$ |
| Fe | | | | | |
| $A$ | $14.286.10^{0}$ | $-14.034.10^{0}$ | $14.170.10^{0}$ | $-46.817.10^{-1}$ | $48.10^{-4}$ |
| $B$ | $10.075.10^{-1}$ | $-16.512.10^{-1}$ | $16.591.10^{-1}$ | $-54.656.10^{-2}$ | $30.10^{-4}$ |
| $\sigma$ | $-31.104.10^{-1}$ | $86.755.10^{-1}$ | $-84.324.10^{-1}$ | $26.525.10^{-1}$ | $29.10^{-4}$ |
| $r_o$ | $-68.043.10^{-1}$ | $17.928.10^{0}$ | $-17.132.10^{0}$ | $53.219.10^{-1}$ | $17.10^{-4}$ |

The calculation of CLLDF was estimated by using the modeled function that represented by Eq.5 with its parameters (Eq.7) at the fixed primary energy $3.10^{15}$ eV around the knee region of CR energy spectram for different primary particles. Fig. 4 shows the results of the simulated CLLDF (solid lines) and that parameterized (dashed lines) of primary particles ($e^-$, n and p) at zenith angle $\theta$=10$^o$. While Fig.5 demonstrates the results of the parametrized CLLDF (dashed lines) in comparison with that simulated with CORSIKA code at zenith angle $\theta$=25$^o$ for different primary particles (F, K and Fe). The parameterized CLLDF in Fig. 4 and Fig. 5 for inclined EAS showers slightly differs from CLLDF simulated for configuration of Yakutsk EAS array. This difference was identified by minimization of the formula:

$$\Delta = \sum_i \left[ \frac{Q_{Par}(\theta, R)}{Q_{COR}(R)} - 1 \right]^2 \rightarrow \min \tag{8}$$

Where the results for each primary particle are given in Table 3.

**Table 3:** Results of $\Delta_{min}$ that calculated by (Eq. 8) for each primary particle at the primary energy 3 PeV and at zenith angles 10° and 25°.

| θ=10° | | |
|---|---|---|
| ***electron*** | ***neutron*** | ***proton*** |
| 16.961.10$^{-6}$ | 33.880.10$^{-4}$ | 14.846.10$^{-3}$ |
| **θ=25°** | | |
| ***Fluorine*** | ***Potassium*** | ***Iron*** |
| 83.940.10$^{-4}$ | 22.676.10$^{-4}$ | 35.669.10$^{-6}$ |

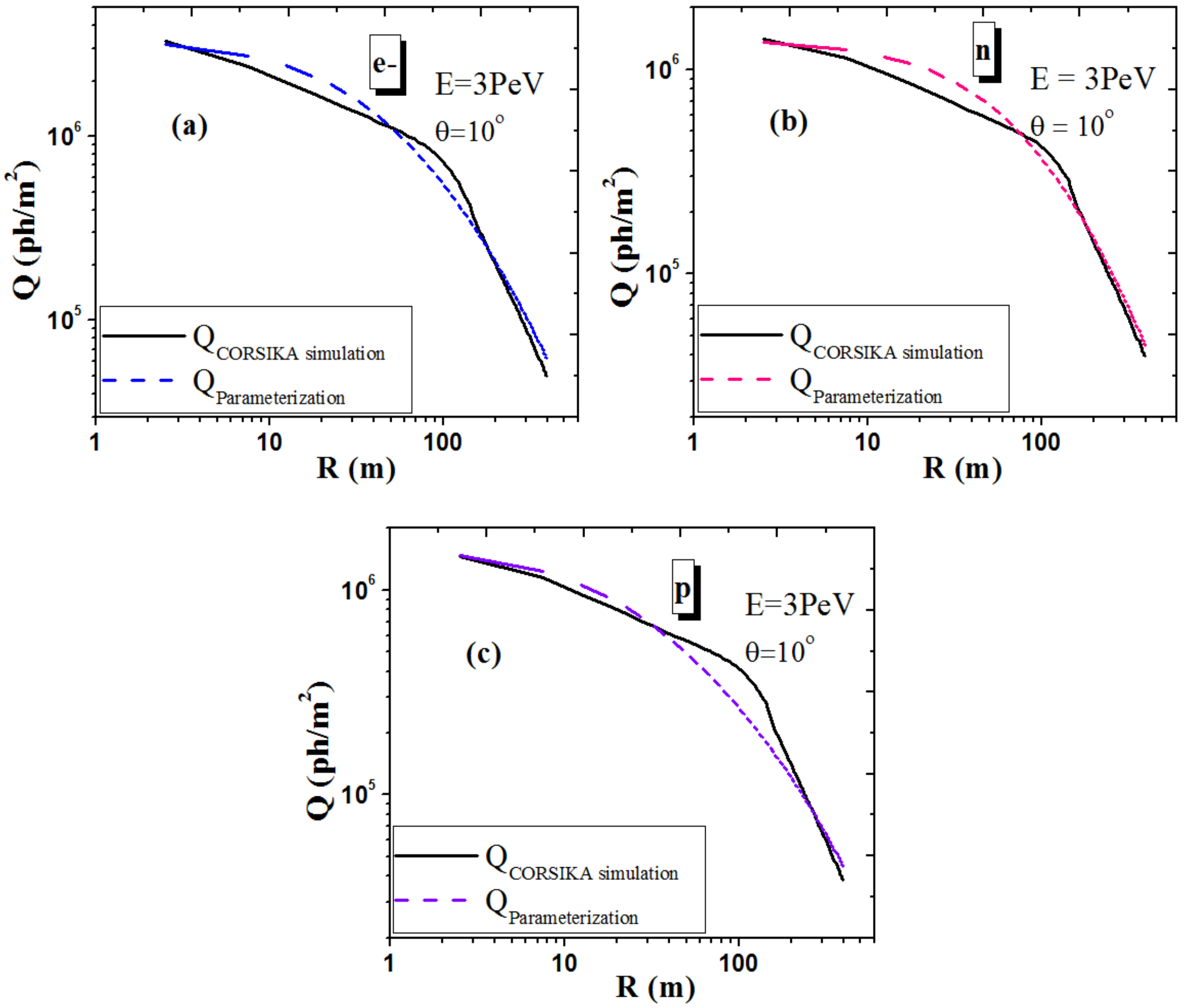

**Fig. 4** Comparison of the simulated CLLDF with CORSIKA code for Yakutsk array conditions with the energy 3 PeV at θ=10° (solid lines) and one approximated using Eq. 5 (dashed lines) for: (a) e$^-$; (b) n and (c) p.

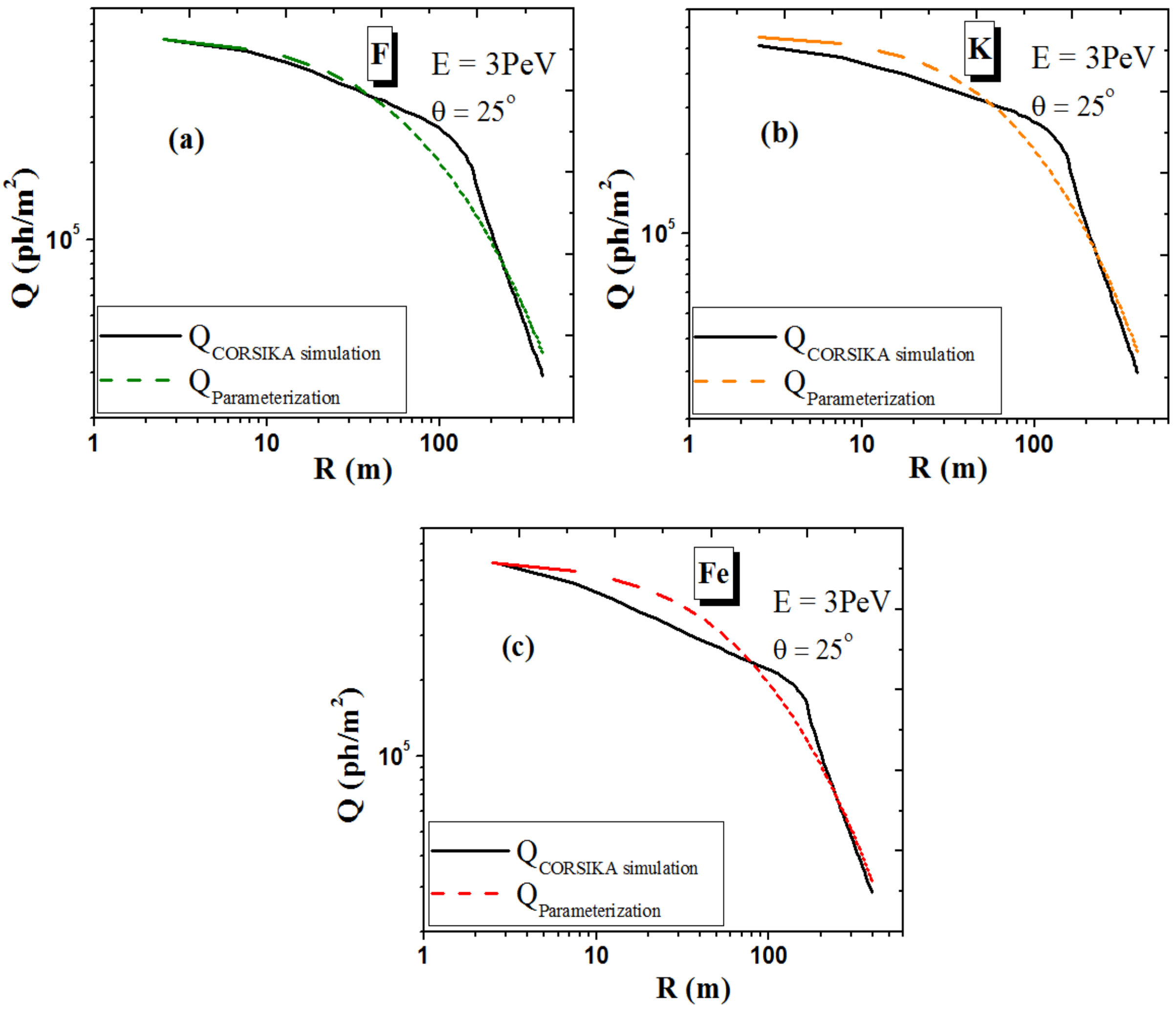

**Fig. 5** Comparison of the simulated CLLDF using CORSIKA code for Yakutsk array configurations at the energy 3 PeV and θ=25° (solid lines) and one approximated using Eq. 5 (dashed lines) for: (a) F; (b) K and (c) Fe.

The parameterized CLLDF (dashed lines) compared with Yakutsk experimental data (symbol lines) are demonstrated in Fig. 6 for the primary particles (p, F and Fe) at zenith angle $\theta$=5°. While Fig.7 shows the results of the parametrized CLLDF (dashed lines) in comparison with the Yakutsk experimental data at zenith angle $\theta$=15° for the primary particles ($e^-$, n and K). This comparison was identified by minimization of the functional:

$$\Delta = \sum_i \left[ \frac{Q_{Par}(\theta, R)}{Q_{\exp}(R)} - 1 \right]^2 \rightarrow \min \qquad (9)$$

Table 4 shows the results of $\Delta_{\min}$ calculated by Eq. 9**.**

**Table 4:** The results of $\Delta_{min}$ that calculated by Eq. 9 for each primary particle at the primary energy 3 PeV at zenith angles $5^o$ and $15^o$.

| **θ=5°** | | |
|---|---|---|
| ***proton*** | ***Fluorine*** | ***Iron*** |
| $59.376.10^{-3}$ | $28.392.10^{-3}$ | $14.364.10^{-3}$ |
| **θ=15°** | | |
| ***Electron*** | ***Neutron*** | ***Potassium*** |
| $43.048.10^{-4}$ | $94.260.10^{-4}$ | $26.195.10^{-4}$ |

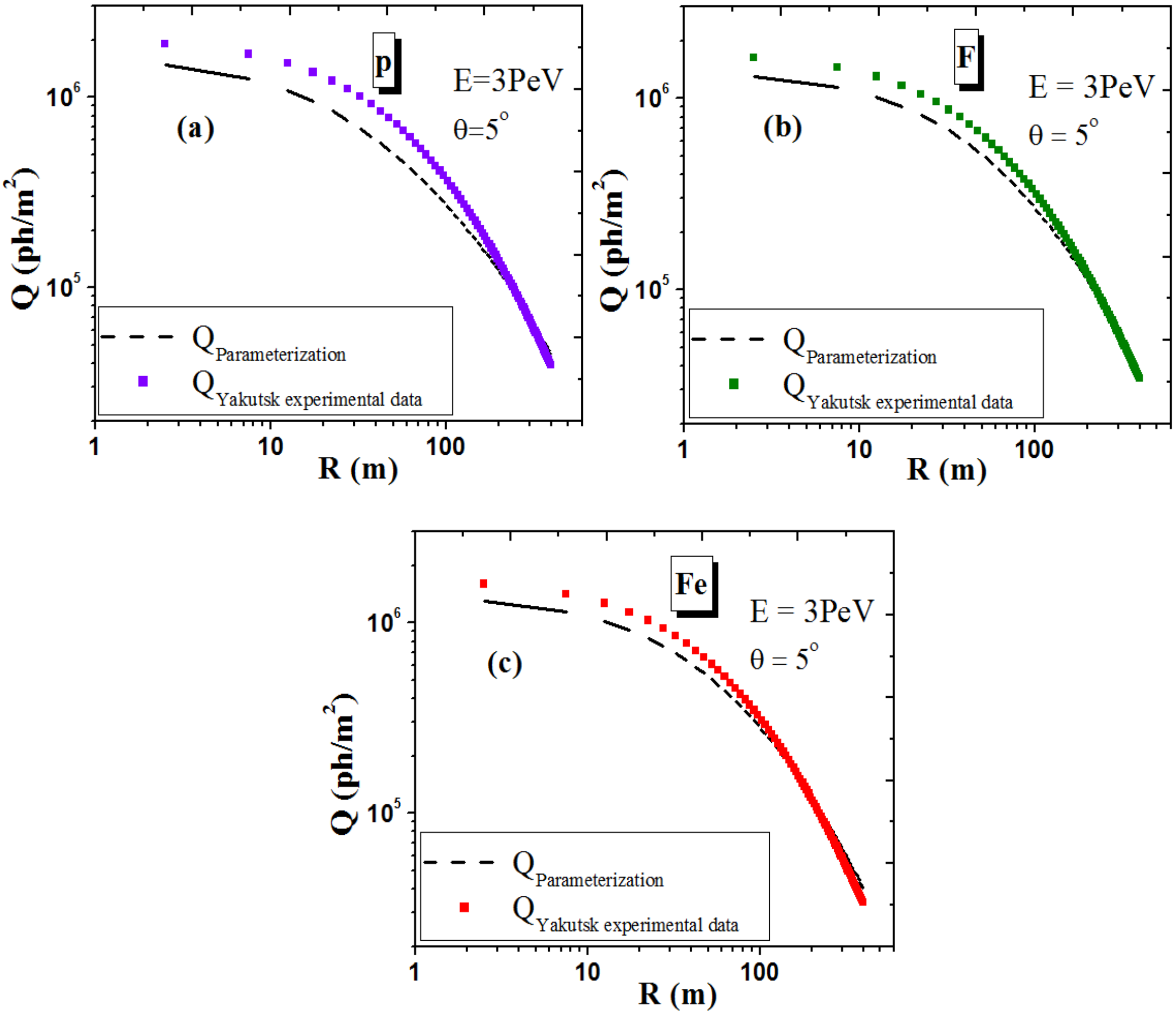

**Fig. 6** Comparison of the parameterized CLLDF using Eq. 5 (dashed lines) with the experimental data obtained by Yakutsk EAS array (symbols) with the energy 3 PeV and $\theta=5^o$ for: (a) p; (b) F and (c) Fe.

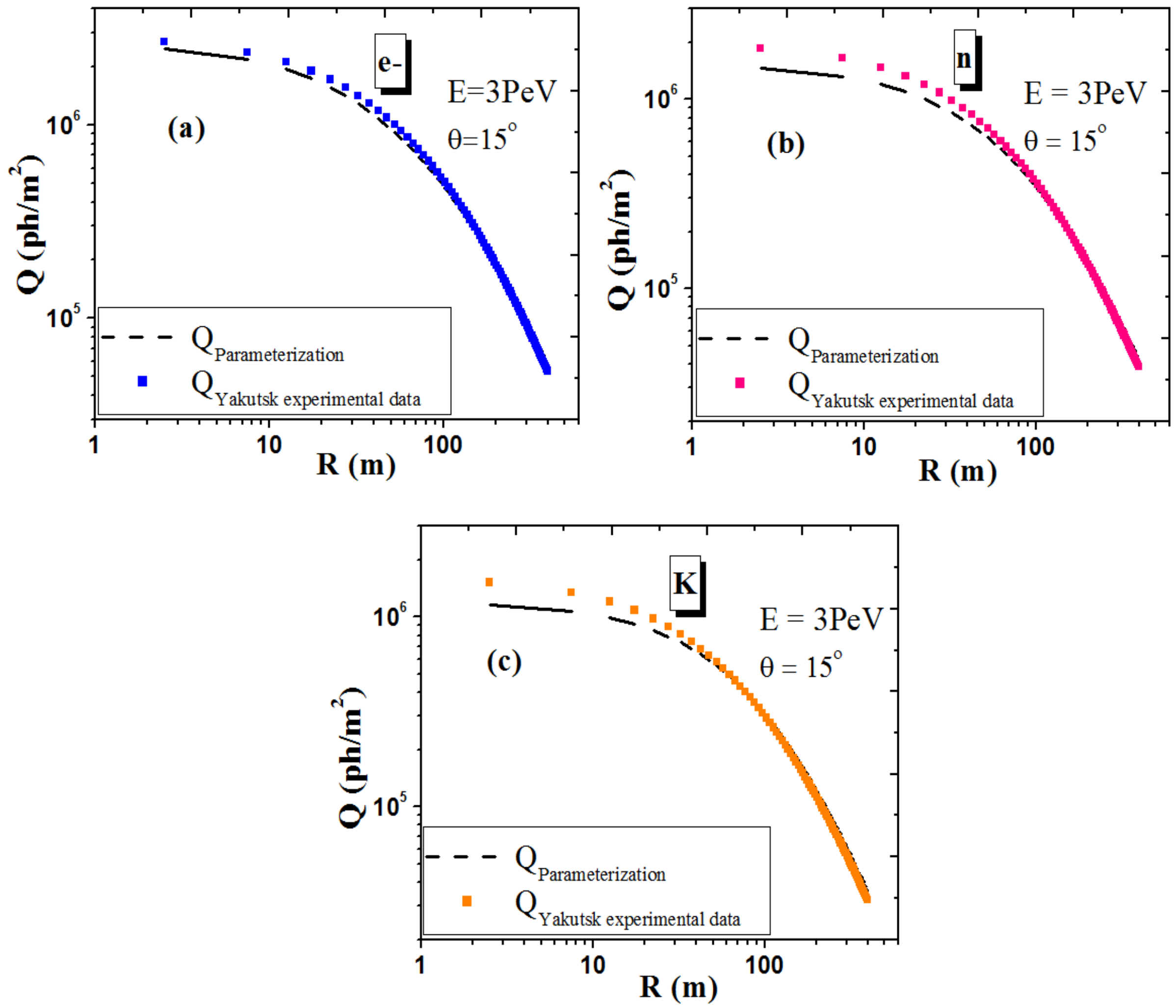

**Fig. 7** Comparison of the parameterized CLLDF using Eq. 5 (dashed lines) with the experimental data obtained by Yakutsk EAS array (symbols) with the energy 3 PeV and $\theta=15^o$ for: (a) $e^-$; (b) n and (c) K.

## 5. Conclusion

The simulation of the Cherenkov light lateral distribution function in extensive air showers was performed for conditions and configurations of the Yakutsk EAS array for different primary particles ($e^-$, n, p, F, K, and Fe) around the knee region with the energy $3.10^{15}$ eV. Basing on Breit-Wigner function, an approach that describing the Cherenkov light lateral function in EAS and analyzing the ability of its application for reconstructing the registered events on the Yakutsk EAS array. A parameterization of CLLDF was reconstructed depending on this simulation as a function of the zenith angle. The estimated CLLDF at zenith angles $10^o$ for ($e^-$, n and p) and $25^o$ for (F, K and Fe) with primary energy 3PeV demonstrated a good agreement with that simulated by CORSIKA code for Yakutsk EAS array. The parameterized CLLDF also compared with that measured with Yakutsk EAS array for ($e^-$, n and K) at zenith angle $15^o$ and for (p, F and Fe) at zenith angle $5^o$. The approximation constructed on the basis of this calculation allowed us to reconstruct preliminary the events, including reconstruction of the type of particle generating the EAS, and to estimate the particle energy from the signal amplitudes of Cherenkov light registered with the Yakutsk EAS facility.